![metals logo] *metals* — MDPI

*Article*

# Symmetry Breakdown Related Fracture in 42CrMo4 Steel

**Jian Feng \*, Stefan Barth and Marc Wettlaufer**

Centre of Materials Engineering, Heilbronn University, Max-Planck-Str. 39, 74081 Heilbronn, Germany; stefan.barth@hs-heilbronn.de (S.B.); marc.wettlaufer@hs-heilbronn.de (M.W.)

\* Correspondence: jian.feng@hs-heilbronn.de

**Abstract:** Austenite grains that underwent the f.c.c. to b.c.c. (or b.c.t.) transformation are typically composed of 24 Kurdjumov–Sachs variants that can be categorized by three axes of Bain transformations; thus, a complete transformation generally displays 3-fold symmetry in (001) pole figures. In the present work, crystallographic symmetry in 42CrMo4 steel austempered below martensite start temperature was investigated with the help of the orientation distribution function (ODF) analysis based on the FEG-SEM/EBSD technique. It is shown that, upon phase transformations, the specimens contained 6-fold symmetry in all (001), (011), and (111) pole figures of an ODF. The ODF analysis, verified by theoretical modeling, showed that under plane-strain conditions cracks prefer to propagate through areas strongly offset by the high symmetry. The origin of high symmetry was investigated, and the mechanism of the symmetry breakdown was explained.

## 1. Introduction

42CrMo4 (AISI 4140, DIN/EN 1.7225) is the most widely used quenched and tempered steel due to its outstanding properties/cost ratio. Some researchers believe that this alloy has been completely understood, while others are trying to investigate this alloy at the atom scale [1] and dedicating themselves to austempering [2,3], coating [4], and 3D-printing [5] of this versatile material. During the hardening of 42CrMo4, austenite grains that undergo the f.c.c. to b.c.c. (or b.c.t.) transformation are typically composed of 24 Kurdjumov–Sachs variants that can be categorized by three axes of Bain transformations [6]; thus, a complete transformation generally displays 3-fold symmetry in (001) pole figures. Kinney et al. [7] reported that 6-fold symmetry could exist in 42CrMo4 steel by analyzing the pole figures. After this important work, limited research on symmetry in 42CrMo4 was performed to the best of the authors' knowledge, and orientation distribution function (ODF) analysis in the nature of a superposition could be a better method to handle this issue.

As described by Biezeno and Grammel in 1953 [8], if the crack is free to propagate along a certain path, it must choose the path which minimizes the Lagrangian function, the stable solutions of which exhibit less symmetry than the equation itself, hence releasing the maximum energy from the system. This is true for dynamic as well as quasi-static crack propagation [9]. In this sense, the fracture could either be a consequence or an origin of spontaneous symmetry breaking, a very common occurrence in physics. In the present consideration, the resistance to crack propagation is vital for industrial applications of ultra-high-strength low-alloy (USLA) steel such as 42CrMo4, heat-treated to have a yield strength over 200 ksi (1379 MPa) [3,10]. After spontaneous phase transformation, high crystallographic symmetry, especially 6-fold orientation symmetry, could always be present in these steels. However, the effect of crystallographic orientation symmetry and/or symmetry breakdown on the overall properties of steel of a particular fracture, which could be significant, is still unknown [7]. To figure out this problem, both experimental and theoretical investigations are performed.

## 2. Materials and Methods

The chemical composition and metallurgical features of the versatile 42CrMo4 steel are represented in Table 1. The cycles of heat treatment, the preparation of circumferentially notched round bars, and the performance of the notched tensile testing have been described somewhere else [11]. Cracks were introduced by the notched tensile testing, interrupted right before the load reached the average notched fracture load. Crystallographic orientations of the microstructure of various specimens were measured at room temperature by the field emission gun scanning electron microscope/electron backscattered diffraction (FEG-SEM/EBSD) technique and analyzed by MatLab-MTEX toolbox. The accelerating voltage for EBSD measurements was 20 kV, the step size was <100 nm and the tolerance range of the Euler-angle measurements is 0.5–1° in these observations. The overall hit rate was >80%, which is approx. 10% lower than the measurements of uncracked specimens. ODF estimation in MTEX was based upon the modified least squares estimator. The recalculated/reconstructed pole figures of an ODF were compared with the measured data to ensure identical crystallographic symmetry.

**Table 1.** Chemical composition (in wt %) of the investigated 42CrMo4 steel; volume fraction of retained austenite ($f_{RA}$, in %), effective grain size ($d_{eff}$, in μm) and aspect ratio ($\kappa$, in -) of lath-shaped units in 42CrMo4 steel austempered below $M_S$. w.a. refers to weighed by area and w.l. weighed by length, respectively.

| C | Si | Mn | S | P | Cr | Mo | Fe | $f_{RA}$ | $d_{eff}$ [1] | | $\kappa$ [2] | |
|---|---|---|---|---|---|---|---|---|---|---|---|---|
| | | | | | | | | | w.a. | w.l. | w.a. | w.l. |
| 0.40 | 0.35 | 0.82 | 0.018 | 0.007 | 0.90 | 0.11 | Bal. | 0 | 2.5 | 0.9 | 0.4 | 0.4 |

[1] Estimated with the maximal misorientation angle of 15°. [2] Peak position, fitted by the Pseudo-Voigt function.

## 3. Results

Figure 1a shows the inverse pole figure (IPF) map of the quasi-homogeneous microstructure on the plane–strain crack path. In the IPF maps, the lath-morphology of the microstructure remained after the recalculation of EBSD maps with the maximal misorientation angle of 15°. Pole figures of an ODF were rotated to show the symmetry around the (111) poles. Area I displays high symmetry in Figure 1c, which compares to the ideal 6-fold symmetry, as represented in Figure 1b. Area II displays strongly deformed 6-fold symmetry (Figure 1d). Notably, strong (110) orientations, indicated by A in Figure 1e, appear on the great circle of the (111) pole figure in the plane-strain cracked Area III and lead to a strong offset of the 6-fold symmetry.

Figure 2 represents the Euler sections to visualize ODF in Area I, II, and III, calculated based on a cubic crystal and monoclinic sample symmetry from the {111} poles. The monoclinic symmetry delivers the most stable and distinguishable texture components. As crystallographic components (E1 and E2) and (J1 and J2) cannot appear independently for monoclinic symmetry; in the analyses and discussion that follow, distinctions will not be made between them. The observed b.c.c. texture components are listed in Table 2. In Figure 2a, the only major orientation evident is J. The remaining major orientations D1, D2, and F are practically absent. The intensity of E is approximately half that of J. In Area II visible are the major orientations D1, E, F, J, and the transformation texture (110) [1–10]. In particular, the intensity of (110) [1–10] compares to that of J, while the intensity of D1 is approximately half that of D2 (Figure 2b). In plane-strain cracked Area III, the intensity of (110) [1–10] is approximately twice that of D2, E, F, and J. The {110} fiber extends from F through J to E along the $\varphi_2$ direction (Figure 2c).

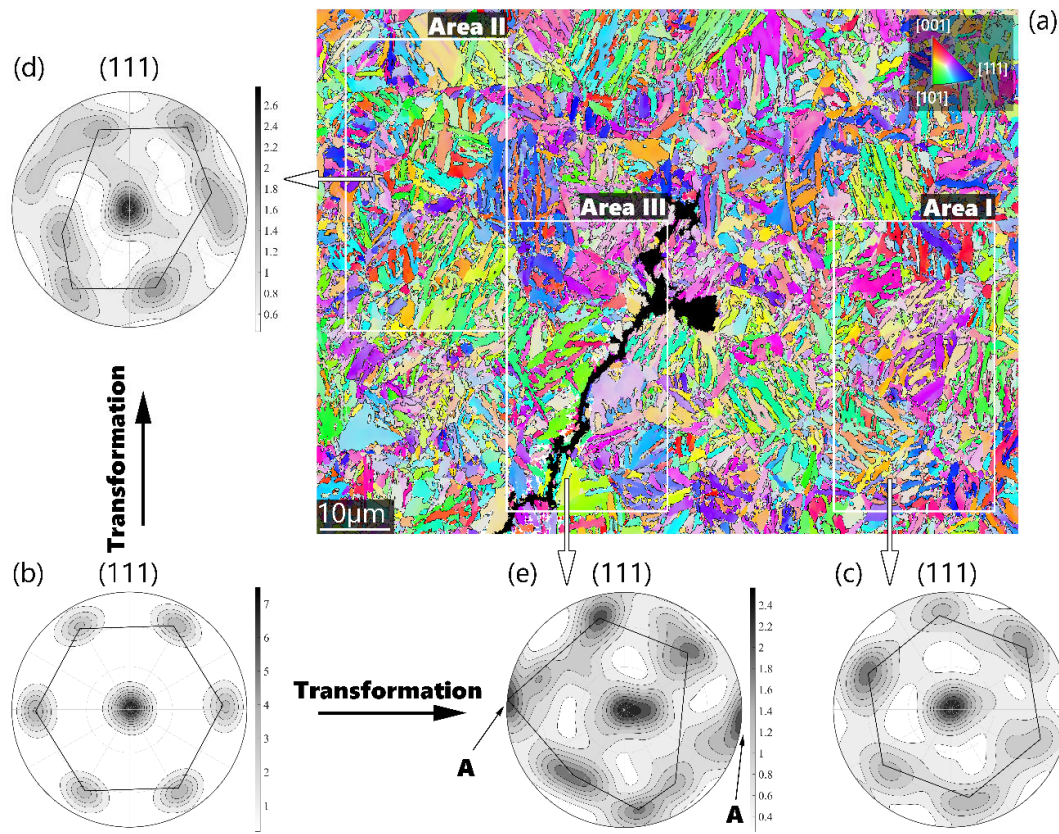

**Figure 1.** Observed symmetry-breakdown-related fracture in steel. (**a**) Inverse pole figure (IPF) coloring of fractured lath-shaped microstructure in 42CrMo4 steel austempered below *M*s. Colors of crystals agree with the orientations perpendicular to the observed plane, which is indicated in the stereographic triangle; (**b**–**e**) Pole figures of an orientation distribution function (ODF) rotated to represent the symmetry about the (111) pole: (**b**) Ideal 6-fold symmetry, (**c**) evident 6-fold symmetry in Area I, (**d**) strongly deformed 6-fold symmetry in Area II, and (**e**) 6-fold symmetry is undetectable in cracked Area III.

Figure 3 shows the grain boundary misorientation distribution of the investigated hierarchical microstructure. The distribution compares to the one for fully lath-martensitic [12] or martensitic/lower bainitic [13] microstructures. The strong peak observed near 60° in the grain boundary misorientation distribution is relevant to the relative orientations between the variants in crystallographic packets of the hierarchical structure. A high fraction of boundaries with a misorientation in the range 2.5–8° were also found to be located inside the martensite laths, typically forming sub-laths [14].

Figure 4 shows the distribution of misorientation angles around the [011] axis between neighboring variants in Area I. Similar misorientation distributions were also obtained for Areas II and III. In the figure, several misorientation peaks, at about 13°, 31°, 41°, 51–55°, 61°, and 71°, are seen. This observation of multiple misorientation peaks suggests that the dominating orientation relationship is not exactly that due to Kurdjumov–Sachs (K–S) [15] or Nishiyama–Wassermann (N–W) [16]. The K–S and N–W relationships are the most commonly observed during displacive transformations, albeit small deviations do exist. As shown in Figure 4, some observed peak misorientations are close to, but not the same as, the values 10.5°, 49.5°, 60°, and 70.5°, expected from the K–S orientation relationship. The low misorientation angle population approximately in the range of 20–40° could represent the misorientation angles inherited from the prior austenite grains [12].

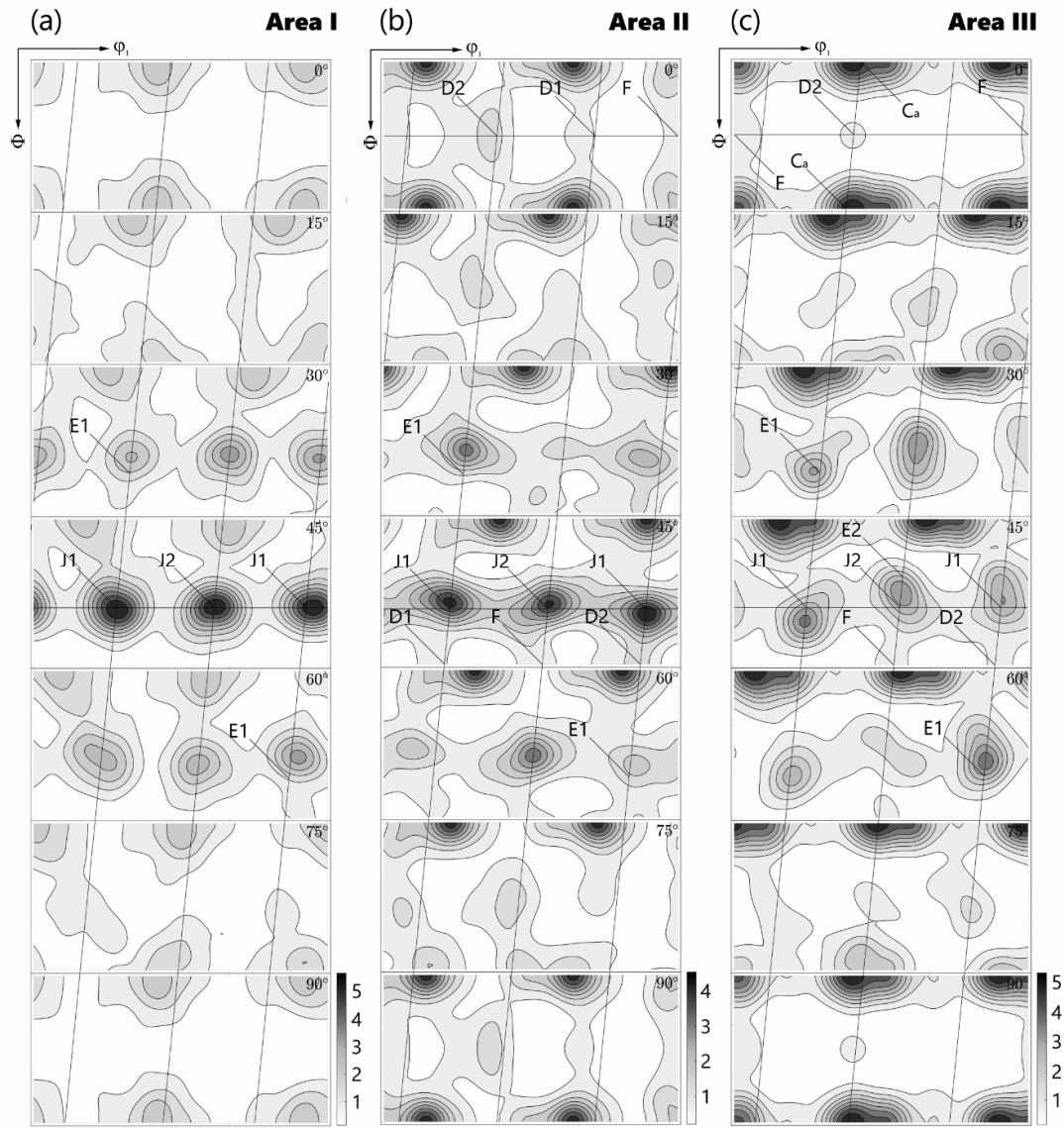

**Figure 2.** Euler sections to visualize ODF in (**a**) Area I, (**b**) Area II, and (**c**) Area III with monoclinic symmetry.

To investigate the origin of high symmetry, IPF coloring of selected α-blocks (with maximal misorientation angle ≤5°) is shown in Figure 5a. The results of the correspondent ODF analysis on the scanned area are shown in Figure 5b to represent the 6-fold symmetry about the (001), (011), and (111) poles. The point orientations from the individual blocks are connected with solid lines in roughly identical colors of blocks. It is shown that each block in Figure 5a (denoted by the number 1, 2 or 3) contributes to almost one-third of the 6-fold-symmetric pattern in the pole figures of an ODF. The repetition or rotation of twinning may explain the observation (see below). Figure 5c shows the fluctuations in misorientations along with the line segments A–B, C–D, and E–F drawn in Figure 5a. The line segment A-B goes across three block boundaries, and these boundaries were determined to be Σ3, i.e., twin boundaries. A boundary relatively close to the Σ1 orientation was also found within the same block. The line segments C-D and E-F lay inside the individual blocks, where only small misorientations ≤5° were found.

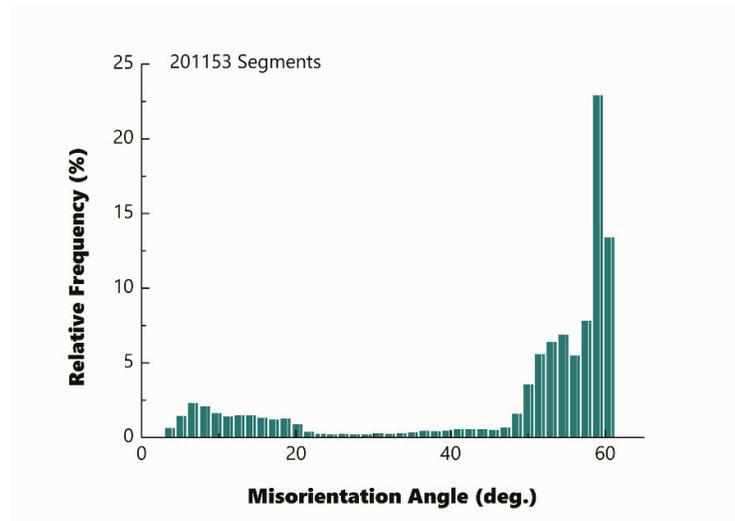

**Figure 3.** Experimental grain boundary misorientation distribution of the hierarchical microstructure investigated. The grains with misorientation ≤2° were treated as one unit.

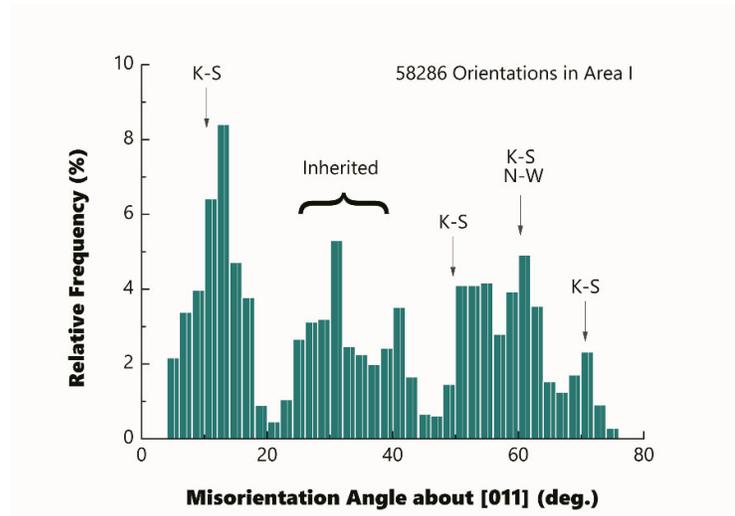

**Figure 4.** Experimental distribution of misorientation angles about [011] axis between neighboring variants in Area I. K–S indicates Kurdjumov–Sachs, and N–W Nishiyama–Wassermann orientation relationships, respectively.

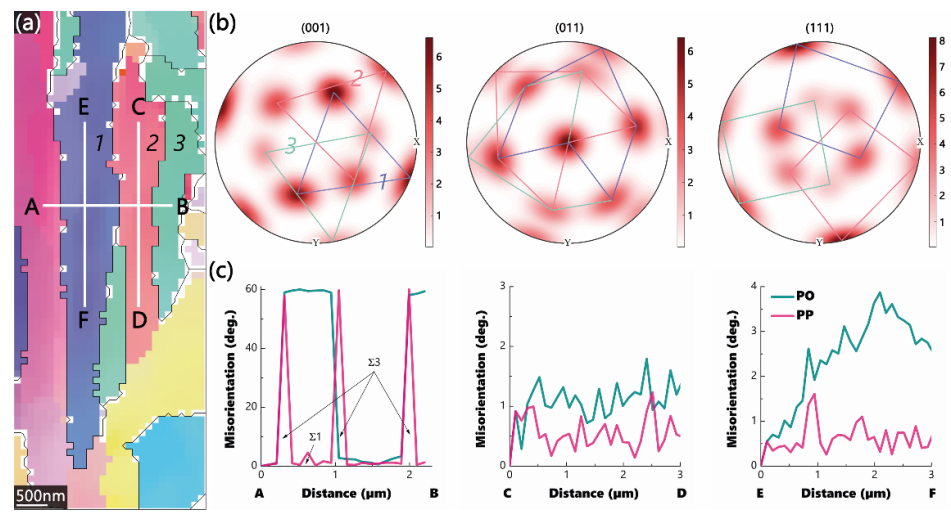

**Figure 5.** Origin of the 6-fold symmetry. (**a**) Inverse pole figure (IPF) coloring of lath-shaped microstructure in 42CrMo4 steel austempered below $M_S$. The pixel (step) size of the EBSD scanning is <100 nm; (**b**) pole figures of an orientation distribution function (ODF) of the scanned area in (**a**) to represent the symmetry about the (001), (011), and (111) poles. The point orientations from the individual blocks in (**a**) are connected with solid lines in the identical color blocks; (**c**) fluctuations of misorientations along the straight lines A–B, C–D, and E–F drawn in (**a**). PO: point-to-origin misorientation; PP: point-to-point misorientation.

## 4. Discussion

Figure 1 indicates that areas with the lowest symmetry in pole figures of an ODF may crack first, whereas areas with higher symmetry can hold in the plane-strain state, under the precondition that the observed break in the symmetry was not developed by the deformation of the b.c.c. mixed microstructure, but rather directly by the transformation of sheared f.c.c. austenite. It needs be verified that the lower symmetry weakens the materials, and it is not the plastic deformation or even failure of the material that breaks the 6-fold symmetry developed by the phase transformation. The observed development of the texture, as shown in Figure 2 implies, that this statement is indeed valid. A switch of the dominant texture components from J to (110) [1–10] has been observed from Area I to III. The (110) [1–10] components have been emphasized in Area II and dominate in Area III, changing the symmetry of these areas significantly. As demonstrated in Table 2, the (110) [1–10] components are developed directly by the transformation of sheared f.c.c. austenite, which was preliminarily analyzed by Wittridge and Jonas [17]. As a result, it can be concluded that the transformation components break the 6-fold symmetry, and, therefore, the plane-strain fracture toughness of the material deteriorates.

**Table 2.** Description of observed b.c.c. texture components developed either directly by shear (S) or by the transformation of sheared austenite (T).

| Symbol | Component | Type | Shear [1] |
|---|---|---|---|
| J1 | (0–11) [–211] | S/T | 0.11–0.58 |
| J2 | (1–10) [–1–12] | S/T | 0.11–0.58 |
| D1 | (11–2) [111] | S/T | 0.07–0.29 |
| D2 | (–1–12) [111] | S/T | 0.94 |
| E1 | (01–1) [111] | S/T | 0.09–0.81 |
| E2 | (0–11) [111] | S/T | 0.09–0.81 |
| F | (110) [001] | S/T | 0.05–0.09 |
| C$_\alpha$ | (110) [1–10] | T | 0.58 |

[1] Dependent on the parent components of the γ-α transformation.

Theoretical modeling of symmetry breakdown related fracture in steel is made under the following assumptions:

A. The crack through the fractured subunits (i.e., blocks or laths) obeys elastic-plastic fracture mechanics (EPFM) and has a plastic zone upon the plane-strain condition;
B. The inter-subunit fracture is insignificant after austempering below the martensite start temperature (ABMS);
C. The radius of the selected integration contour for the path-dependent J-integral is quantized by the average effective path of the crack propagating through individual subunits;
D. A uniform distribution of the crystallographic symmetry of grains is assumed.

The boundary of the plastic zone is a function of $\theta$, i.e., the angle between the radius of the plastic zone and the propagating direction of the crack tip. For a plane-strain case, it is well known [18]

$$r(\theta) = \frac{1}{4\pi}\left(\frac{K_I}{\sigma_y}\right)^2 \left[(1-2\nu)^2(1+\cos\theta) + \frac{3}{2}\sin^2\theta\right] \quad (1)$$

where $\sigma_y$ is the yield strength, $K_I$ the stress intensity factor, and $\nu$ the Poisson's ratio. Obviously,

$$r(\theta) \leq \frac{\eta}{4\pi}\left(\frac{K_I}{\sigma_y}\right)^2 \quad (2)$$

i.e.,

$$r_y = \frac{\eta}{4\pi}\left(\frac{K_{IC}}{\sigma_y}\right)^2 \quad (3)$$

with the maximum of the non-dimensional $r(\theta)/(K_I/\sigma_y)^2$

$$\eta = \frac{3}{2} + (1-2\nu)^2 + \frac{1}{6}(1-2\nu)^4 \quad (4)$$

Beyond the maximal radius of the plastic zone $r_y$, the selected integration contour for J-integral is greater than the plastic zone size and J-integral becomes path-independent.

After assumption C, it yields

$$r_y = f\bar{L} \quad (5)$$

where $f$ is the frequency of crack propagation, say, $f = \phi/c$, for qualitative discussions. Here $\phi$ is an experimentally accessible and material-dependent length parameter: for instance, the grain size. Upon assumption D, it is then possible to calculate the average crack propagation length $\bar{L}$ on a 2D-intersection of a 3D-$\alpha$ (martensitic/bainitic) ellipsoid in the form of an ellipse of major-axis $a$ and minor-axis $c$, with $a \gg c$ and $\kappa = c/a$. $\bar{L}$ can be simultaneously treated as the quantized energy needed to cut through ellipsoids and calculated from the (N-1)th root of the (N-1)th products of chord lengths of an ellipse

$$\bar{L}^{N-1} = \frac{N}{c}\left[\left(\frac{a+c}{2}\right)^N - \left(\frac{a-c}{2}\right)^N\right] = \frac{N}{2^N c}[(a+c)^N - (a-c)^N] \quad (6)$$

Here, $\bar{L}^{N-1}$ is $N/c$ times the Nth Fibonacci number. As illustrated in Figure 6, $N$ is called the geometric number, controlled by the transformation symmetry of the substructure with $N \in \mathbb{N}$ and $N \gg 2$.

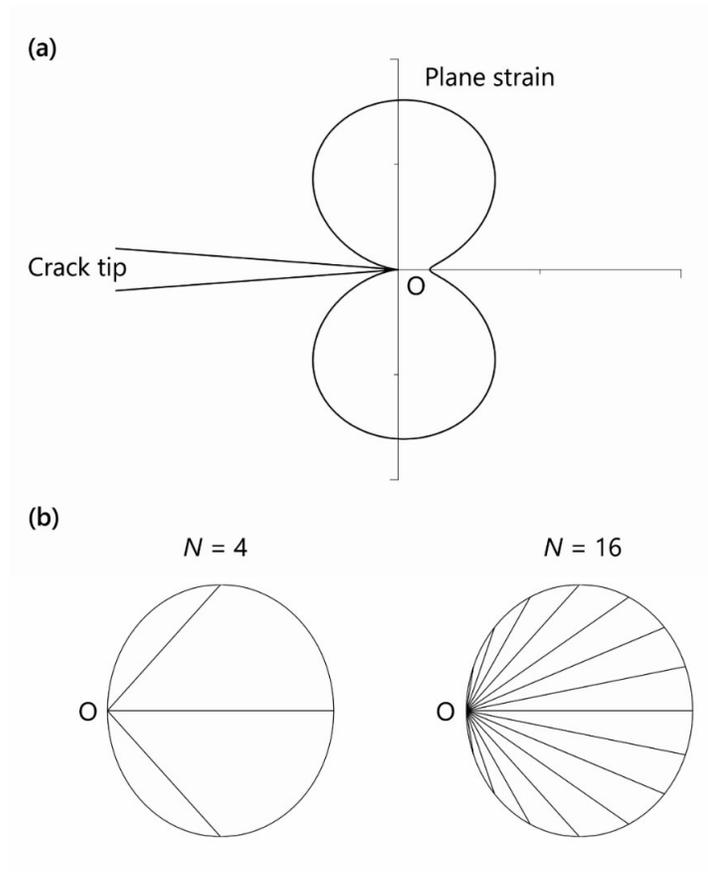

**Figure 6.** (**a**) Plastic zone shapes in Mode I according to Von Mises yield criteria. The boundary of the plastic zone is plotted non-dimensionally according to Equation (1). O is the location where the crack tip is about to propagate. (**b**) Schematic explanation of the geometric number *N*. The individual chord length of a 2D-intersection of a 3D-α ellipsoid is denoted by *L* and the average chord length by $\bar{L}$, respectively. *N-1* chords cut the intersection equally. The schemata have a superposition nature and combine the possible pathways of the crack. High symmetry generally diminishes *N* and enlarges $\bar{L}$. According to Equation (7), $K_{IC,N=4} > K_{IC,N=16}$.

Combining Equations (3) and (6) yields

$$K_{IC} = \sigma_y \omega \sqrt{\frac{2\pi\phi}{\eta}} \tag{7}$$

with

$$\omega = \left\{\frac{N}{2}\left[\left(\frac{1}{\kappa}+1\right)^N - \left(\frac{1}{\kappa}-1\right)^N\right]\right\}^{\frac{1}{2(N-1)}} \tag{8}$$

After the binomial expansion, Equation (7) turns to

$$K_{IC} = \sigma_y \sqrt{\frac{2\pi\phi}{\eta}} \left[N \sum_{k=1}^{m} \binom{N}{2k-1} \frac{1}{\kappa^{N-(2k-1)}}\right]^{\frac{1}{2(N-1)}} \tag{9}$$

with $\exists m \in \mathbb{N}: (N+1)/2 = m$.

Evaluating the limit of Equation (9) yields

$$K_{IC,Sym} > \lim_{N \to \infty} K_{IC} \tag{10}$$

with $\sigma_{y,Sym} \geq \sigma_{y,N\to\infty}$ owing to the strengthening effect of the high symmetry. $K_{IC,Sym}$ is the critical stress intensity factor, if $N$ is small (say, $N$ is against 2). Upon constant $\kappa$, $\omega$ increases monotonically with decreasing $N$. If $N$ is small (but still $\gg 2$), only certain energetic pathways of the crack, which are constrained/controlled by the symmetric equivalent orientations, can be selected (as shown in Figure 6b). On the contrary, as $N$ approaches infinity, the propagation of the crack becomes unconstrained and the pathways of the crack across the subunits become random. The modeling shows that constraints induced by geometric symmetry can enlarge $\bar{L}$, and hence enhance the fracture toughness.

As shown in Figure 5, the origin of high symmetry could be the repetition and/or rotation of the twinning. Here, the repetition and/or rotation operations should be statistically and carefully arranged, otherwise 6-fold symmetry cannot be constructed by the point orientations with significant intensity. The symmetry breakdown could be closely related to f.c.c. to b.c.c. (or b.c.t.) phase transformation. It is known that high symmetry in prior austenite grains comes from the thermal twins (which are basically mechanical in nature). After phase transformation, this high symmetry is inherited by $\alpha$-constituents (bainite/martensite). Although the simplest type of displacive transformation involves lattice changes that are closely similar to those in deformation twinning [19], the difference between these changes is indeed obvious. In displacive transformation, the strain direction is no longer confined within the invariant plane; therefore, if the magnitude of the strain is large enough, shear deformation may break the symmetry (or the order), which was established by the arranged twinning.

## 5. Conclusions

In summary, ODF analysis with IPF maps was used to account for the crystallographic symmetry breakdown in 42CrMo4 steel austempered below $M_S$. Fracture toughness and crystallographic orientation symmetry were connected by EPFM. The results of experimental studies, explained by the modeling, show that symmetry breakdown deteriorates $K_{IC}$ and/or facilitates fracture. The origin of high symmetry was investigated and the mechanism of the symmetry breakdown was explained. The present work shows a remnant of a deep symmetry in the laws of the material that are hidden from sight, and could be of universal significance for understanding the fracture behavior of steel transformed at low temperatures.

**Author Contributions:** conceptualization, methodology, investigation, J.F.; data curation, J.F., S.B. and M.W.; writing—original draft preparation, J.F.; writing—review and editing, S.B. and M.W. All authors have read and agreed to the published version of the manuscript.

**Funding:** Open Access funding enabled and organized by Prorektorat Forschung, Transfer, Innovation, Heilbronn University, Germany

**Data Availability Statement:** Not applicable.

**Acknowledgments:** Dedicated to the memory of S. Bührer.

**Conflicts of Interest:** The authors declare no conflict of interest.